\documentclass{aamas2012}

\usepackage{graphicx}
\usepackage{amsmath}
\usepackage{amssymb}
\usepackage{float}
\usepackage{url}

\newtheorem{definition}{Definition}

\def\predi{\Gamma^\pi_i}

\def\AP{\mathcal S_n}

\title{Crowd IQ - Aggregating Opinions to Boost Performance}

\begin{document}

\numberofauthors{5}

\author{
\alignauthor
Yoram Bachrach \\
Microsoft Research \\
yobach@microsoft.com \\
\alignauthor
Thore Graepel \\
Microsoft Research \\
thoreg@microsoft.com \\
\alignauthor
Gjergji Kasneci \\
Microsoft Research \\
gkasneci@gmail.com \\
\and 
\alignauthor
Michal Kosinski \\
University of Cambridge \\
mk583@cam.ac.uk \\
\alignauthor
Jurgen Van Gael \\
Microsoft Research \\
jurgen.vangael@gmail.com \\
}

\maketitle

\begin{abstract}
We show how the quality of decisions based on the aggregated opinions of the crowd can be conveniently studied using a sample of individual responses to a standard IQ questionnaire. We aggregated the responses to the IQ questionnaire using simple majority voting and a machine learning approach based on a probabilistic graphical model. The score for the aggregated questionnaire, Crowd IQ, serves as a quality measure of decisions based on aggregating opinions, which also allows quantifying individual and crowd performance on the same scale.

We show that Crowd IQ grows quickly with the size of the crowd but saturates, and that for small homogeneous crowds the Crowd IQ significantly exceeds the IQ of even their most intelligent member. We investigate alternative ways of aggregating the responses and the impact of the aggregation method on the resulting Crowd IQ. We also discuss Contextual IQ, a method of quantifying the individual participant's contribution to the Crowd IQ based on the Shapley value from cooperative game theory.
\end{abstract}

\category{I.2.11}{Artificial Intelligence}{Distributed Artificial Intelligence}[Multiagent Systems]

\terms{Algorithms, Economics}

\keywords{Human Intelligence, Opinion Aggregation, Shapley Value}


\section{Introduction}

Human intelligence and group decision processes have been extensively studied for many years. However, in recent years internet-based technologies have dramatically changed the ways in which people interact, socialize and communicate. People exchange information through online social networks, communicate their opinions through websites and rely on internet sources in their economic decisions.
The accessibility of such information makes it easier to \emph{aggregate} the opinions of many individuals and examine the quality of decisions based on such aggregated information.

While \emph{collecting} the opinions of individuals is easy, it is more difficult to decide how to \emph{aggregate} the opinions in order to \emph{reach decisions}, and \emph{measure the quality} of the decisions based on the aggregated information. There is often a considerable variance in the individual opinions so reaching optimal decisions is not a trivial task. Moreover, once a decision has been reached, it is impossible to compare its quality to other possible decisions that have not been taken. Consequently, although aggregating the opinions of many individuals is appealing, performing the aggregation process successfully requires answering many important questions: How does the quality of the decisions depend on the group size? Is it better to base decisions on the opinions of the best individual in a group, or is it better to rely on other people's opinions as well? How can one measure the contribution of individuals to the quality of the decisions? Do individual contributions depend more on the individual's skill or on her similarity to the group?

In an attempt to answer those questions we explore the process of aggregating participants' responses to IQ items on the established IQ test\footnote{We refer to individuals who have completed an IQ test as \emph{participants}, questions in such a test as \emph{items}, and to participants' answers as \emph{responses}.} - Raven's Standard Progressive Matrices (SPM)~\cite{raven1978manual}. We treat participant' responses to an IQ test, expressed independently in a setting similar to popular crowdsourcing environments such as Amazon Mechanical Turk, as their opinions regarding the correct solution. We aggregate those individual opinions using majority vote or a machine learning aggregator to reach a decision regarding a correct response to each of the items. We then score this test solved by the crowd using standard scoring procedures, referring to the resulting IQ score as \emph{the Crowd IQ}.
 
People's responses to the SPM IQ test offer a convenient and robust environment to study the aggregation of individual opinions. First, SPM offers a set of non-trivial problems engaging a range of human cognitive abilities with well-defined correct response and limited number of possible solutions. Second, an individual's IQ score is an elegant measure of one's mental abilities and is a good predictor of behavior and performance in a broad spectrum of contexts including job and academic performance, creativity, health-related behaviors and social outcomes~\cite{gottfredson1997g,jensen1998g,lubinski2004introduction,schmidt2004general}. Third, the Crowd IQ score provides a convenient quality measure of the crowd's aggregated decision. Finally, IQ scores provide a uniform performance metric that allows exploring the relationship between individual and crowd performance.

{\bf Our Contribution:} We examine the properties of the Crowd IQ and show that aggregating opinions of crowd members can significantly boost the expected quality of the decision. We show that the Crowd IQ grows quickly with its size but then saturates, indicating diminishing returns from each additional member. We also show that for homogeneous crowds the Crowd IQ significantly exceeds the IQ of the most intelligent member in the crowd. Finally, we show that an individual's contribution to the Crowd IQ is not solely related to the participant's IQ but also depends on the uniqueness of her contribution in the context of a given crowd.

\section{Related Work}
\label{l_related_work}

Many papers deal with aggregating the opinions of multiple agents to reach high quality decisions. Social choice theory deals with joint decision making by self-interested agents (see~\cite{sen1986social} for a broad discussion of this field), and is a key research area in artificial intelligence and multiagent systems. The Condorcet Jury Theorem from social choice theorey provides theoretical bounds regarding the probability of a set of agents to reach the correct decision under majority voting~\cite{austen1996information,mclennan1998consequences,list2001epistemic}. However, the Condorcet Jury Thereom uses strong assumptions which may not hold in practice, such as requiring votes to be completely independent. Our study can be viewed as an empirical examination of this topic using data from IQ questionnaires.

Another related field is judgment aggregation~\cite{list2009judgment} which deals with aggregating group members' individual judgments on some interconnected propositions, expressed in a formal logic language, into corresponding collective judgments on these propositions. These fields have also been examined by computer scientists, who found practical and computationally tractable ways of performing such aggregation, ranging from machine learning approaches~\cite{kasneci2010bayesian} to prediction markets~\cite{pennock2007computational}. However, the IQ test items are not interconnected, and our focus is on quantifying and decomposing the group's performance. In our paper we also ignore the complications of aggregating agent opinions when such agents are self-interested and may wish to influence the aggregated choice~\cite{everaere2007strategy,dekel2008incentive}.

Human intelligence has been a central topic in psychology. Psychologists noted that people's performance on many cognitive tasks is strongly correlated, leading to the emergence of a single statistical factor, typically called ``general intelligence''~\cite{spearman1927abilities,gottfredson1997g,lubinski2004introduction,schmidt2004general}. Recent work extends this to ``collective intelligence'' for performance of \emph{groups} of people in joint tasks~\cite{woolley2010evidence}, which is not strongly correlated with the maximal or average intelligence of the group members. However, this approach examines explicit collaboration and interaction between the group members, where the social interaction may sometimes even hinder performance~\cite{lorenz2011social}, whereas we focus on information aggregation. Approaches more similar to ours are~\cite{dawid1979maximum,lyle2008collective} and~\cite{welinder2010multidimensional} which even proposes a machine learning aggregator for image labeling in a crowdsourcing environment. However, our focus is on the impact of the aggregation methods and methods for quantifying individual contribution based on a standardized IQ test. 

\section{Methodology}
\label{l_methodology}

We now describe the IQ test we used, our dataset of participants' responses, and the aggregation methods used to establish the crowd's solution to the test. 

\subsection{Standard Raven Progressive Matrices Test}
\label{l_sect_raven}

\begin{figure}[ht!]
 \centering
 \includegraphics[width=0.48\textwidth,height=0.35\textheight]{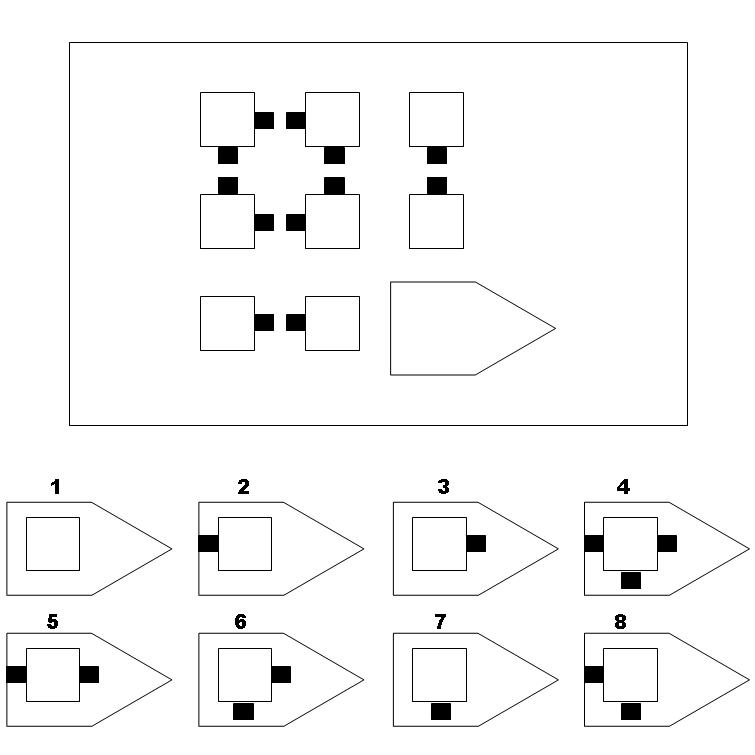}
 \caption{Item similar to those in the SPM test}
 \label{fig:RavenItem}
\end{figure}

The IQ test used in this study, Raven's Standard Progressive Matrices~\cite{raven:spm:2008,raven2000raven}, was
developed by John C. Raven~\cite{raven1938progressive}. It is a multiple choice non-verbal intelligence test drawing on Spearman's theory of general ability~\cite{spearman1927abilities} and consists of $m = 60$ matrices with one element missing and $k = 8$ possible responses. Matrices are separated into five sets of 12 and within each set the problems become increasingly difficult. A sample item, similar\footnote{The SPM test is copyright protected, so we can only provide an item similar to those in the actual test, rather than a sample item from the test itself.} to those used in the SPM is shown on Figure~\ref{fig:RavenItem}. 

Raven's SPM and its other forms (Advanced and Colored Progressive Matrices) are one of the most popular intelligence tests used in both research and clinical settings, as well as in high-stake contexts such as in military personnel selection and court cases~\cite{raven2000raven}. 

\subsection{Dataset and Scoring}
\label{l_sect_data_collection}

Our sample consisted of 138 individuals, aged 15-17, who filled the SPM during its standardization for the British market in the year 2006~\cite{raven:spm:2008}. 
The sample is representative of the British population. 

\begin{figure}[ht!]
 \centering
 \includegraphics[width=0.5\textwidth,height=0.25\textheight]{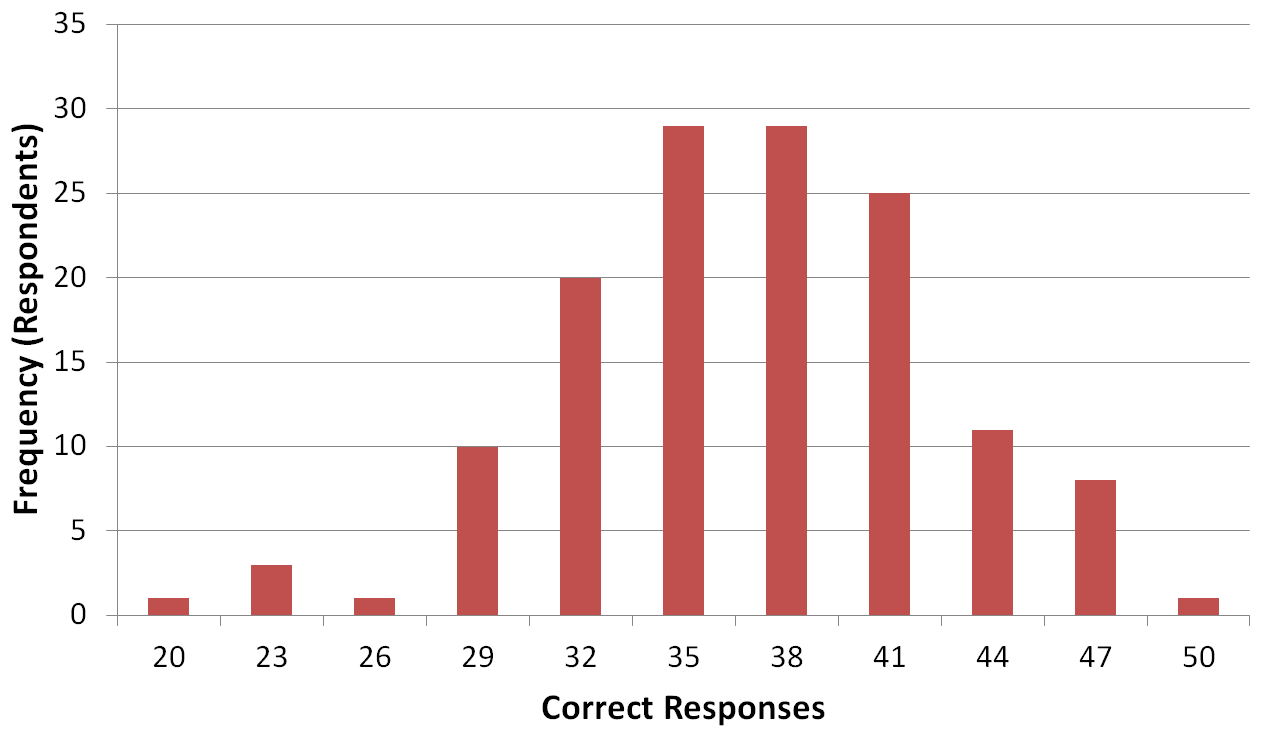}
 \caption{Histogram of raw IQ scores}
\label{rawhist}
\end{figure}

\begin{figure}[ht!]
 \centering
 \includegraphics[width=0.5\textwidth,height=0.25\textheight]{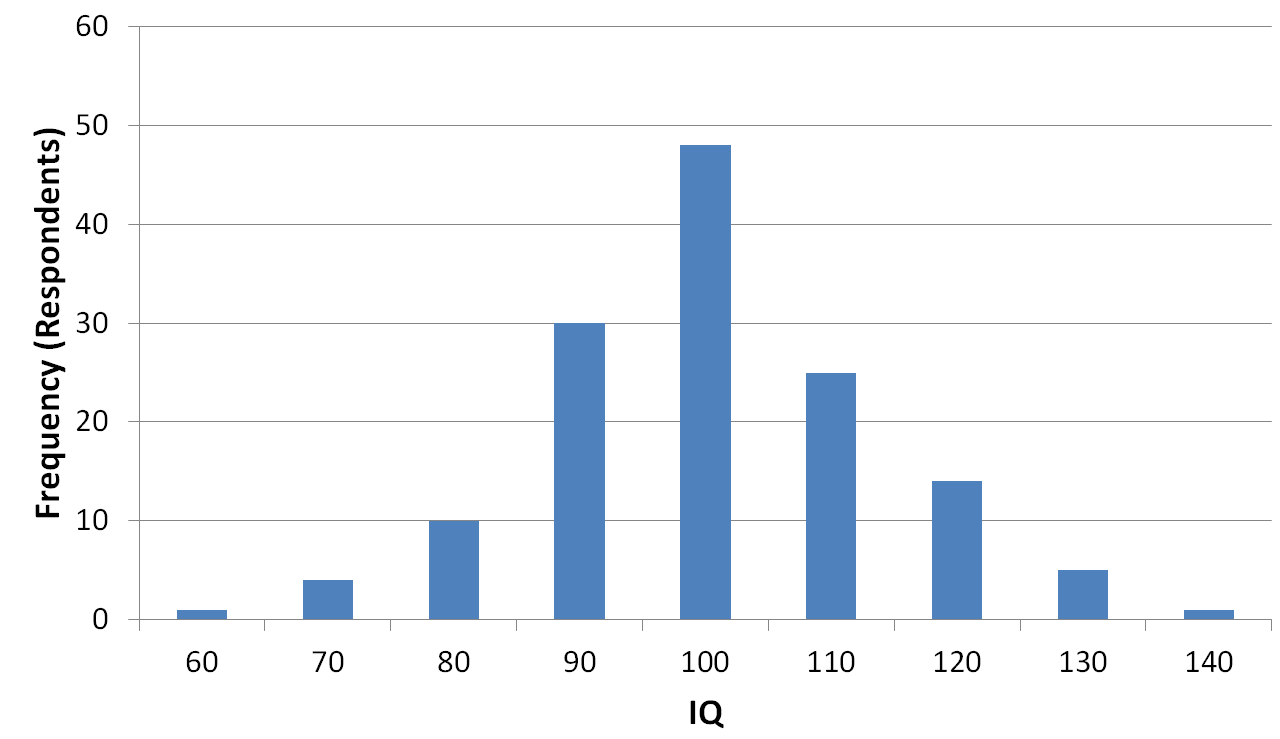}
\caption{Histogram of IQ scores}
\label{iqhist}
 \end{figure}

The standard scoring procedure described in the test's manual was used to calculate individual and Crowd IQ scores~\cite{raven:spm:2008}. The manual provides tables for translating the number of correct responses (raw score) into an IQ score. The IQ scale characteristic for SPM (and most other intelligence tests) is standardized on a representative population to have a normal distribution with an average score of 100 and standard deviation of 15. Hence, IQ scores allow for convenient comparisons between individuals, and comparing individual performance with the general population. The distribution of the raw scores and the IQ scores in our sample is shown on Figures~\ref{rawhist} and~\ref{iqhist}. The average number of correct responses in the dataset is 36.04, with a standard deviation of 5.49. The average IQ score is 99.57 with a standard deviation of 14.16.

\subsection{Aggregating Individual Responses}
\label{l_sect_agg_iq}

Consider an IQ questionnaire consisting of $m$ items, each a multiple choice item with $k$ possible responses. Denote the possible responses $K=\{1,\ldots,k\}$. The questionnaire is administered to a set $N$ of $n$ participants, each providing a response for each of the items. Let $r^i_j \in K$ be the response provided by participant $j \in N$ to item $i$ and $r_j$ be the responses provided by participant $j \in N$ to all items, so $r_j = (r^1_j, r^2_j, \ldots, r^m_j)$. We call $r_j$ the \emph{filled questionnaire} for participant $j$. 

An aggregation method $f$ takes the filled questionnaires of the participants, $r_1,\ldots,r_n$, and outputs a single filled questionnaire $agg^f_N$, which contains a response to each of the items, so $agg^f_N = (r^1_{agg}, \ldots, r^1_{agg})$ where $r^i_{agg} \in K$ is the response chosen to the item $i$. The aggregated questionnaire $agg^f_N$ is scored using the standard scoring key. 

We now briefly describe the two aggregators used in this paper: \emph{A simple majority aggregator with lexicographical tie-breaking} ({\bf MAJ}), and \emph{a machine learning aggregator} ({\bf ML}). 

\subsubsection{Simple Majority Aggregation}
\label{subsect:simple majority}

The MAJ aggregator considers each item of the IQ questionnaire separately. It chooses the most common response as the ``correct'' response, and thus bases the decision on the choice made by the majority of the participants. If two or more responses are selected an equal number of times (tie) the first one in the lexicographical order is selected. 

The MAJ aggregator has several limitations. First, it does not use the information obtained from the responses to one of the items to decide how to aggregate the responses to another item. For example, if a user $u$ has answered all items correctly until item $i$ and user $v$ has answered all items incorrectly until item $i$, when aggregating the responses to an item $i+1$, it might be desirable to give $u$'s opinion more weight than $v$'s opinion. Further, the MAJ aggregator makes no assumptions about the data-generating process other than that the correct response should be chosen more frequently than any of the incorrect ones. 

\subsubsection{Machine Learning Based Aggregation} 
\label{l_sect_machine_learning_aggregation}

Our ML aggregator addresses the MAJ's limitations. Similarly to the MAJ aggregator, the goal of the ML one is to take questionnaires completed by several participants and output a single questionnaire with inferred correct responses. In contrast to the MAJ aggregator, this is a non-simple aggregation, in which the inferred response to an item also depends on responses provided to other items. The model attempts to make better inferences about the correct responses to items by jointly modeling the participants' aptitude and the correct responses. The underlying assumption is that each participant has an associated probability of knowing the correct response to an item, their aptitude, and that they will randomly guess the answer if they do not know the correct response. The ML aggregator designed for this study employs probabilistic graphical models~\cite{Pearl:88,Koller+Friedman:09}. 

{\bf Probabilistic Graphical Models} allow structurally describing the generative process assumed to underlie the observed data in terms of latent and observed random variables. In the context of Crowd IQ, information like the correct response to an item or the intelligence of a participant would be modeled as unknown latent variables whereas the given response to an item by a user would be an observed variable. The structure of the model is then determined by the conditional independence assumptions made about the variables in the model. Pearl~\cite{Pearl:88} introduced Bayesian Networks 
to encode assumptions of conditional independence in the form of a graph whose nodes represent the variables and whose edges describe the dependencies between variables. We use the more general notion of a factor graph, see e.g.~\cite{Koller+Friedman:09}, to describe the factorial structure of the assumed joint probability distribution among the variables. Once the structure of the model is defined in terms of a factor graph, observed variables can be set to their observed values. Then approximate message passing algorithms~\cite{Koller+Friedman:09} can infer marginal probability distributions of unknown variables of interest such as the correct response to an item or the intelligence of a participant.

\begin{figure}[ht!]
 \includegraphics[width=0.5\textwidth,height=0.25\textheight]{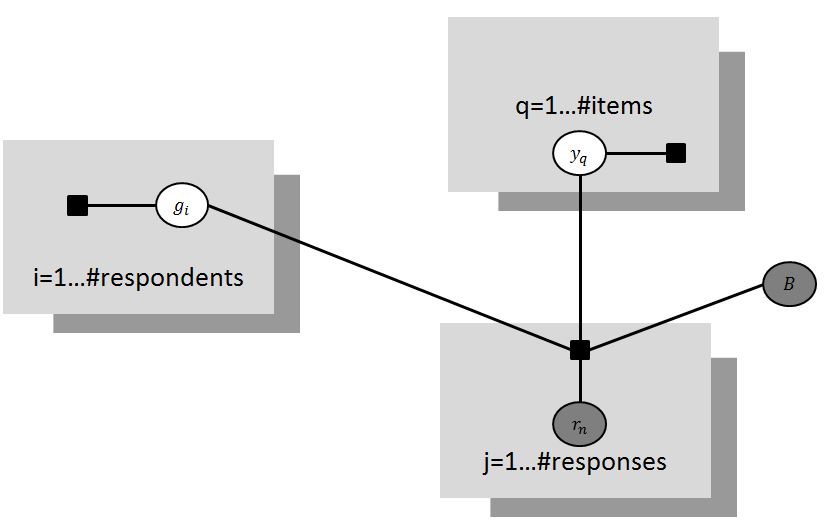}
 \caption{Factor graph for the ML aggregator}
 \label{fig:graphical-model}
\end{figure}

{\bf Graphical Model for IQ Response Data:} We wish to infer the correct responses, so the graphical model contains a set of random variables $y_q \in Y_q$ that represent the correct response to each of the items $q$. Each $y_q$ takes discrete values in the set $Y_q$ of possible responses $q$. The model's initial ignorance about the correct response is expressed by assuming a uniform prior distribution over responses, $y_q \sim \textrm{Uniform}$. We also wish to take into account the (unknown) aptitude of participants in order to weigh their responses appropriately. The aptitude of each participant $i \in N$ is represented by a random variable $g_i \in R$, which can be interpreted as the probability that the participant would know the correct response. We choose uniform prior densities for these variables, $g_i \sim \textrm{Beta}(1.0,1.0)$. Here, $\textrm{Beta}$ represents the family of beta distribution, which allows us to compactly represent (unimodal) beliefs over the $g_i$. We also introduce a uniform ``guessing'' distribution $B = \textrm{Uniform}$, which models the choice of response when the participant is assumed to be guessing. Participant $i$'s response $r_j$ to item $q$, is assumed to be drawn from the following distribution:
\[
 r_j \left\{
 \begin{array}{l l}
 \sim B & \quad \text{with probability} \quad (1 - g_i)\\
 = y_q & \quad \text{with probability} \quad g_i\\
 \end{array} \right.
\]
This means that with probability $g_i$ participant $i$ chooses the correct response $y_q$ and with probability $1-g_i$ she randomly guesses the answer based on the guessing distribution $B$. Figure~\ref{fig:graphical-model} illustrates this probabilistic graphical model in the form of a factor graph. Note that the grey boxes represent plates which indicate repetition of the contained sub-structure of the graphical model. In this case, $q$ ranges over the available items, $i$ ranges over the available participants, and $j$ ranges over the available responses.

Inference in the model is performed using approximate message passing (see \cite{Koller+Friedman:09} for details)\footnote{Our implementation used the Infer.net library. For details regarding Infer.net see: http://research.microsoft.com/en-us/um/cambridge/projects/infernet/).}. As a result we obtain a discrete marginal posterior distribution over responses to each of the items, representing the model's belief about the correct response in light of the observed data. As a by-product we also obtain the posterior marginal densities over the aptitude variables $g_i$ for each user. To minimize the probability of error we choose the response with the maximum posterior for each item as the aggregated response.

\subsection{Contextual IQ: Individual's Contribution to the Crowd IQ}
\label{l_sect_contextual_iq}

We now discuss our approach for quantifying an individual's contribution to a Crowd IQ. Intuitively, individuals of high IQ are likely to contribute more towards the aggregate IQ of the crowd, i.e. the individual's IQ divided by the sum of the IQ scores of all the members of the crowd. However, as individuals' skills and knowledge may differ, the individual contribution depends also on the relationship between the patterns of her responses and those of the other members of the crowd (or context). 
For example, imagine a crowd that can correctly solve a subset of questions $A$ but is unable to provide a correct answer to questions $B$. Adding another individual to this crowd that can correctly solve questions $A$ but does not know correct responses to questions $B$ would not increase the Crowd IQ score, while adding an agent that knows correct responses to questions $B$ can potentially boost Crowd's performance. We refer to this relative boost as a \emph{Contextual IQ}. Our approach to quantifying contextual IQ is based on the Shapley value~\cite{shapley53value}, a concept from cooperative game theory. 

\subsubsection{Measuring Impact on Performance Using the Shapley Value}
\label{l_sect_crowd_iq_through_shapley}

Cooperative game theory studies the behavior of selfish agents who must cooperate to achieve a goal, and analyzes how the rewards from such cooperation should be distributed among the agents. Solution concepts from game theory can be used to find reward distributions fulfilling desirable properties, such as being fair or stable.
Our methodology examines the game where the agents are the participants filling the IQ test, and where the value of a coalition of agents is the Crowd IQ of that coalition. 


The Shapley value~\cite{shapley53value} 
can be viewed as a ``power index'', a tool for measuring an individual's \emph{contribution} or \emph{importance} in the success of a team of agents, or for quantifying an agent's ability to influence a game's outcome~\cite{shapley_shubik1954,banzhaf1965}. The Shapley value was used for measuring political influence of parties forming a coalition in legislative bodies~\cite{Leech:2002:IMF}, analyzing network reliabilitiy~\cite{bachrach2008power,aziz2009power,bachrach2009power} and fair cost allocation~\cite{samet1984application,tijs1986game}. 
Further, the Shapley value is the only imputation fulfilling certain fairness axioms~\cite{shapley53value}.

The Shapley value relies on the marginal contribution of an individual --- the amount of additional utility gained when that individual joins the crowd. We denote by $\pi\in \AP$ a permutation of the agents, so $\pi:\{1,\ldots,n\} \rightarrow \{1,\ldots,n\}$ and $\pi$ is onto. Denote by $\predi$ the predecessors of $i$ in $\pi$, so $\predi = \{j | \pi(j) < \pi(i)\}$. Agent~$i$'s marginal contribution in the permutation $\pi$ is $m^{\pi}_i = v(\predi \cup \{i\}) - v(\predi )$. The Shapley value of an individual is her marginal contribution averaged across all possible permutations of the individuals.

\begin{definition}
\label{l_def_shapley_value}
The Shapley value is the imputation \\ $(\phi_1(v), \ldots , \phi_n(v))$ where
$$\phi_i(v) = \frac{1}{n!} \sum_{\pi \in \AP} m^{\pi}_i = \frac{1}{n!} \sum_{\pi \in \AP} \left( v \left(\predi \cup \{i\} \right) - v \left( \predi \right) \right)$$
\end{definition}

Consider a set $N$ of agents (participants) filling an IQ questionnaire, with $m$ items and a set $K$ of $k$ possible responses to each item, and a set $C \subseteq N$ to be used as a crowd (coalition). Denote the responses of participant $i \in C$ as $r_i \in K^m$, and the set of responses of all the agents in $C$ as $r_C = (r_1, \ldots, r_{|C|})$. Thus the space of possible responses of each agent is $A = K^m$, and the responses of \emph{all} the participants are in the space $A^{|C|}$. Consider an aggregator $f : A^{|C|} \rightarrow A$ which maps the responses of all agents to a single filled questionnaire.

As in Section~\ref{l_sect_agg_iq}, we denote the filled questionnaire obtained by applying the aggregator $f$ to the responses of the agents in $C$ as $agg^f_C$. In Section~\ref{l_sect_agg_iq} we defined the aggregate IQ of a crowd $C$ as the IQ score of the filled questionnaire $agg^f_C$. We define a cooperative game $v_f$ that maps any subset $C \subseteq N$ of agents into their aggregate Crowd IQ (the IQ score of $agg^f_C$, the aggregate response for the crowd $C$). This cooperative game over the set $N$ of agents is defined with the following characteristic function: $v_f(C) = IQ(agg^f_C)$, and is called the \emph{Aggregate IQ Game}.

In the Aggregate IQ Game, the ``reward'' of any coalition $C$ is the aggregate IQ of the crowd $C$, and $v_f(N)$ is the aggregate IQ of the entire agent set $N$. Our goal is to decompose $v_f(N) = IQ(agg^f_N)$, the total aggregate IQ score obtained by the grand coalition $N$ of all agents, to the individual contribution of each agent. We refer to the set $N$ of all agents as the \emph{context} in which we measure an agent's individual contribution. We are thus seeking a vector $\vec p= (p_1,\ldots ,p_n)$ such that $\sum_{i=1}^n p_i = v_f(N)$ where $p_i$ reflects $i$'s fair contribution to the total IQ score. Due to the properties of the Shapley value we can use it to fairly decompose the Crowd IQ score. We define agent $i$'s \emph{Contextual IQ} (for the given context $N$) as its Shapley value in the above Aggregate IQ game. One interpretation of this definition is that the aggregate IQ of the crowd is decomposed into the contribution, in IQ points, of each participant. These contextual IQ scores sum up to the total aggregate IQ of the crowd $N$, and a participant has a higher contextual IQ than another participant if she is expected to have a higher positive influence on the aggregate Crowd IQ score of a subset of participants selected at random from the entire crowd $N$.

By Definition~\ref{l_def_shapley_value}, the contextual IQ is the expected increase in Crowd IQ when adding $i$ to her predecessors in a random permutation of the agent set $N$. Note that $m^{\pi}_i$ is the increase in Crowd IQ when adding $i$ to a specific agent subset, $\predi$, and the Shapley value is the \emph{average} of these increments in Crowd IQ across all agent permutations. Obviously, an individual's contextual IQ (Shapley value) is strongly affected by her IQ score, as responding correctly to more items increases the marginal contribution for $m^{\pi}_i$ for many permutations $\pi$ (assuming a reasonable aggregator).

Computing contextual IQ using formula~\ref{l_def_shapley_value} requires a running time exponential in the number of the agents. We used the approach of~\cite{bachrach2010approximating} for computing the Shapley value, which offers a very high accuracy and a tractable polynomial running time. This algorithm samples many agent subsets (or more precisely permutations) of the crowd and averages the marginal contribution of the target agent in them to obtain an accurate approximation of the Shapley value.

\section{Crowd Size and Crowd IQ}
\label{l_sec_agg_iq_crowd_size}

\begin{figure}[ht!]
 \includegraphics[width=0.5\textwidth,height=0.25\textheight]{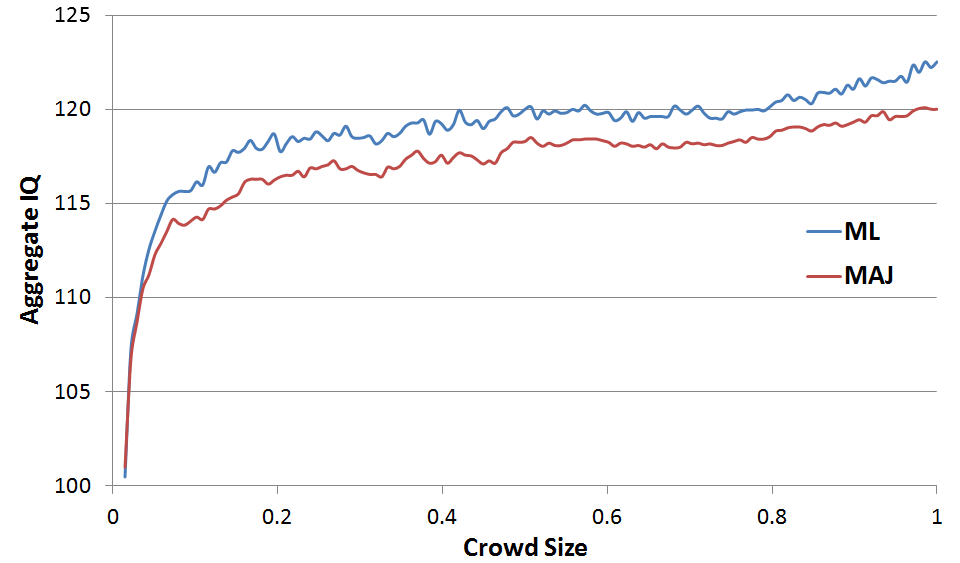}
 \caption{Crowd IQ scores based on the MAJ and ML aggregators for different crowd sizes}
 \label{fig:agg-IQ-MAJ-vs-ML}
\end{figure}

First, we unveil the relationship between the Crowd IQ and its size. Figure~\ref{fig:agg-IQ-MAJ-vs-ML} shows the relationship between the size of the crowd (number of participants) and its IQ established using both MAJ and ML aggregators as discussed in Section~\ref{l_sect_agg_iq}. 
Each point in the plot is the average Crowd IQ for $q=300$ randomly selected crowds of the specified size. Such repetitive sampling minimizes the influence of the selection bias on the Crowd IQ estimates.

Figure~\ref{fig:agg-IQ-MAJ-vs-ML} shows that the Crowd IQ quickly increases with the crowd size but saturates after reaching the crowd size of 14 participants and IQ of about 115, roughly one standard deviation increase above the population mean. These results indicate that the quality of the crowd decision is significantly higher than the average IQ of its members. However, returns from increasing the crowd size rapidly diminish after a certain size is reached. Also, Figure~\ref{fig:agg-IQ-MAJ-vs-ML} shows that a machine learning based aggregation consistently outperforms simple majority aggregation, by learning which users provide correct responses more reliably. 

\section{Smarter Than a Crowd?}
\label{l_sect_agg_iq properties}

Here we investigate whether it is better to base decisions solely on the opinions of the high-performing individuals in a group, or to rely on other peoples' opinions as well. One way of examining this is to determine whether the Crowd IQ is likely to exceed the IQ of the smartest individual in the crowd. We use the approach described in Section~\ref{l_sec_agg_iq_crowd_size} to compute the relationship between Crowd IQ and its size and we plot the maximal individual IQ for any given crowd size. 

\begin{figure}[ht!]
 \centering
 \includegraphics[width=0.5\textwidth,height=0.25\textheight]{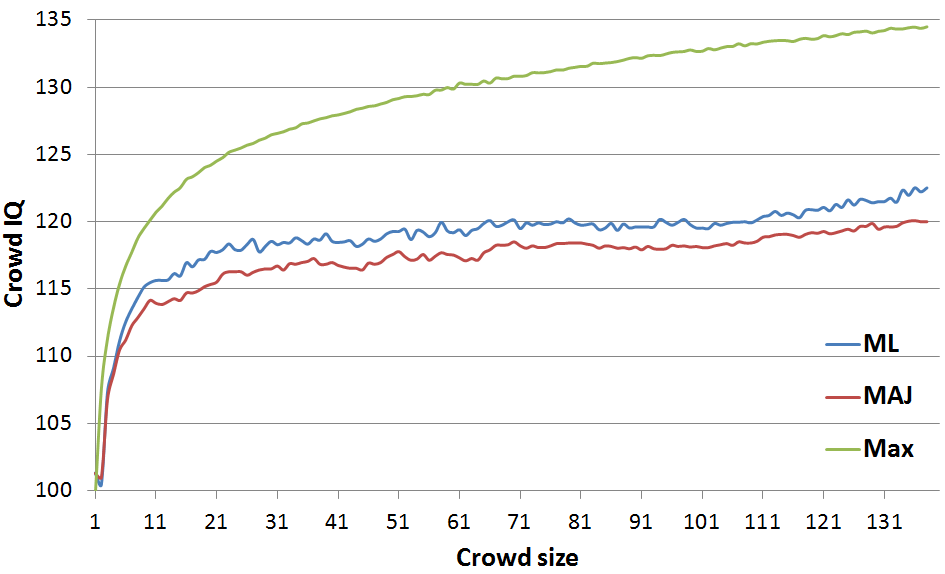}
 \caption{Crowd IQ and maximal IQ (entire dataset)}
 \label{l_fig_agg_iq_vs_max_iq_all_pop}
\end{figure}

Figure~\ref{l_fig_agg_iq_vs_max_iq_all_pop} shows the relation between the crowd size, expected Crowd IQ, and expected maximal IQ for the entire dataset used in this study. It is clear that in large crowds characterized by a wide distribution of IQ scores, the maximal IQ consistently exceeds Crowd IQ. While Crowd IQ for this population saturates around 115-120 IQ points, the chance of the crowd encompassing individuals with extreme IQ scores increases with the sample size. 

However, it is common for the crowds to be composed of individuals characterized by the similar IQ (\emph{homogeneous} crowds). For instance, the IQ of students of advanced degrees is likely to be homogenous and relatively high,as IQ is correlated with academic performance. A homogeneous crowd is less likely to contain an individual with an IQ score much higher than the average IQ score in the crowd, so the performance of the crowd may be superior to that of its smartest individual. To examine this issue, we split our sample into a set of homogeneous subsamples based on the individual IQ scores. Subsamples are denoted by $P_{[L,H]}$, where $[L,H]$ represents the range of participants' IQ scores. Thus, subsample $P_{[110,120]}$ contains individuals with IQ scores between 110 and 120. 

\begin{figure}[ht!]
 \centering
 \includegraphics[width=0.5\textwidth,height=0.25\textheight]{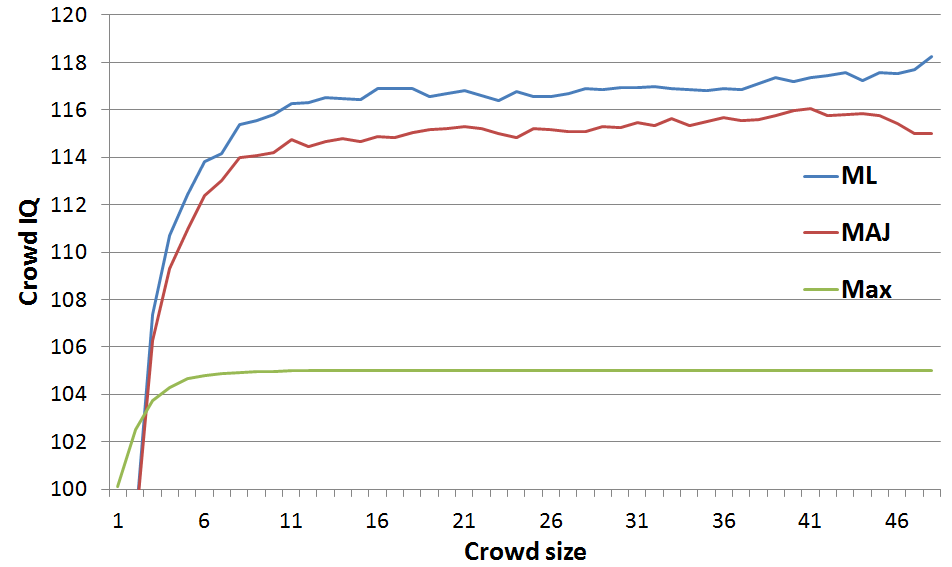}
 \caption{Crowd IQ and maximal IQ for $P_{[95,105]}$ }
 \label{l_fig_agg_iq_vs_max_iq_95_105}
\end{figure}

\begin{figure}[ht!]
 \centering
 \includegraphics[width=0.5\textwidth,height=0.25\textheight]{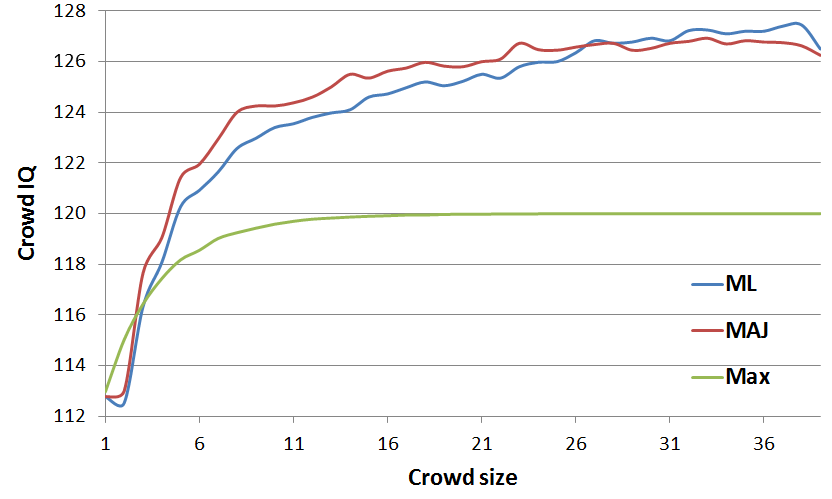}
 \caption{Crowd IQ and maximal IQ for $P_{[110,120]}$ }
 \label{l_fig_agg_iq_vs_max_iq_110_120}
\end{figure}

\begin{figure}[ht!]
 \centering
 \includegraphics[width=0.5\textwidth,height=0.25\textheight]{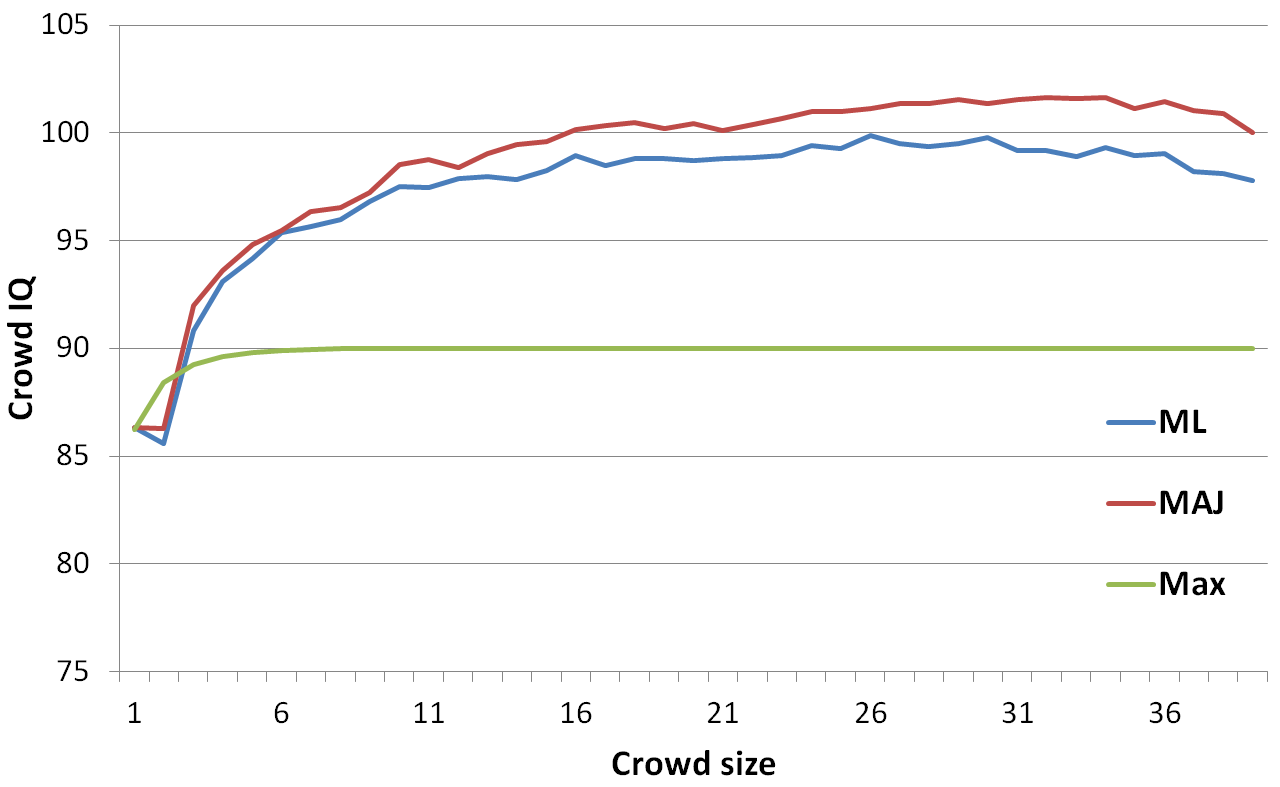}
 \caption{Crowd IQ and maximal IQ for $P_{[80,90]}$ }
 \label{l_fig_agg_iq_vs_max_iq_80_90}
\end{figure}

Figures~\ref{l_fig_agg_iq_vs_max_iq_95_105},~\ref{l_fig_agg_iq_vs_max_iq_110_120}, and~\ref{l_fig_agg_iq_vs_max_iq_80_90} for subsamples $P_{[95,105]}$, $P_{[110,120]}$, and $P_{[80,90]}$\footnote{The number of participants in these subsamples are: $|P_{[95,105]}| = 48$, $|P_{[110,120]}| = 39$, $|P_{[80,90]}| = 39$.} show that the Crowd IQ greatly exceeds its most intelligent member's IQ in homogeneous crowds. Also, a homogeneous crowd's advantage over its smartest member increases as it grows. 

Interestingly, the simple MAJ aggregator outperforms the ML aggregator's in homogeneous high and low IQ populations ($P_{[110,120]}$ and $P_{[80,90]}$). A possible explanation of this phenomenon might be related to the lack of outstanding individuals in such crowds, that could be used by ML aggregator to boost its performance. However, this clearly does not apply to the similarly homogeneous $P_{[95,105]}$ subsample where ML outperforms MAJ aggregator. Further, Figure~\ref{l_fig_agg_iq_vs_max_iq_80_90} shows the decrease in performance of both aggregators for very big crowd sizes. Aggregated performance might be affected by especially popular but incorrect responses to the difficult IQ items that may, for larger crowds, suppress the correct but unpopular responses.

These results indicate that decisions based on the aggregate opinions of rather homogeneous crowds are of a higher quality than those based solely on the opinion of their most intelligent member. On the contrary, in populations characterized by a wide range of individual performance levels, smartest members outperform the crowd. Note, however, that in all cases the Crowd IQ greatly exceeds the IQ of the average member of the crowd, as discussed in Section~\ref{l_sec_agg_iq_crowd_size}. 

\section{Individual IQ and Contextual IQ}
\label{l_indiv_contxt_iq}

\begin{figure}[ht!]
 \centering
 \includegraphics[width=0.48\textwidth,height=0.25\textheight]{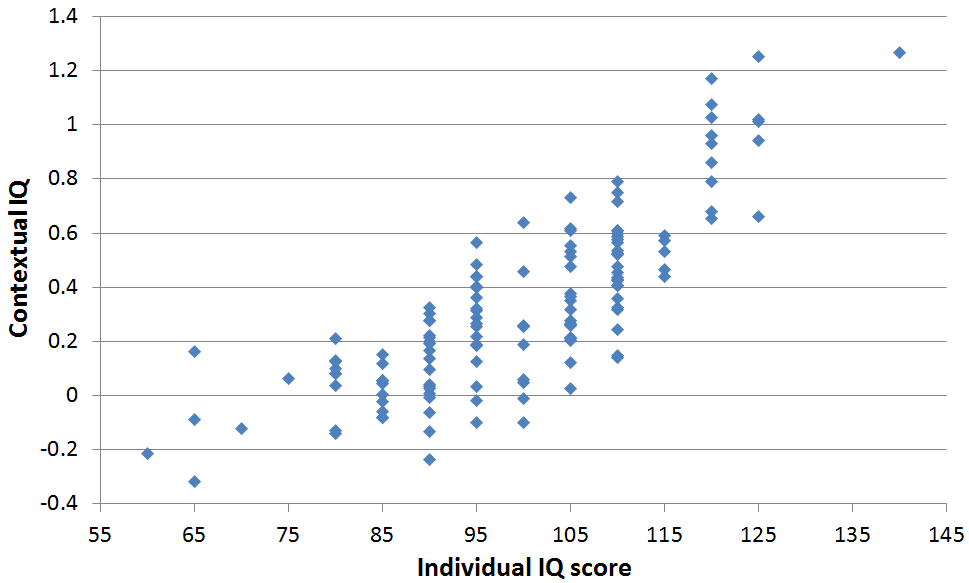}
 \caption{IQ and Contextual IQ}
 \label{l_fig_context_iq_full_ds}
\end{figure}

We now focus on the relationship between individual and contextual IQ. A participant's contextual IQ is the expected increase in Crowd IQ from adding that participant to a random permutation of the crowd's members. Given the opportunity to add another member to a team of an unknown composition, the optimal choice is the agent with the highest contextual IQ.
We now discuss the correlation between individual IQ and contextual IQ using the crowd composed of the entire population of $n=138$ participants. Figure~\ref{l_fig_context_iq_full_ds} presents a scatter plot correlating the participants' IQ scores with their contextual IQ scores. 



As Figure~\ref{l_fig_context_iq_full_ds} shows, there is a positive correlation between IQ and Contextual IQ, but also a high variance of contextual IQs for participants of equal IQ. For example, for the above-average IQ of 105, contextual IQ ranges from very high, through negligible to negative. Thus, even if more intelligent people are generally contributing more to the Crowd IQ, the value of their contribution varies and may even be negative. This indicates that although adding the participant with highest IQ score is a good heuristic, better results can be achieved by using the Contextual IQ approach.

A participant's contextual IQ depends on the aggregated IQ, which in turn depends on the aggregator used. The cooperative game used to generate Figure~\ref{l_fig_context_iq_full_ds} was based on the MAJ aggregator. Measuring the Crowd IQ under a different aggregator (e.g. the ML aggregator), changes the contextual IQ scores of the participants. For example, as shown in Figure~\ref{fig:agg-IQ-MAJ-vs-ML}, the Crowd IQ of all the participants is slightly higher under the ML aggregator, and as the contextual IQ scores must sum up to the total Crowd IQ, the sum of the contextual IQ scores under the ML aggregator would be slightly higher than their sum under the MAJ aggregator. 

\begin{figure}[ht!]
 \centering
 \includegraphics[width=0.5\textwidth,height=0.25\textheight]{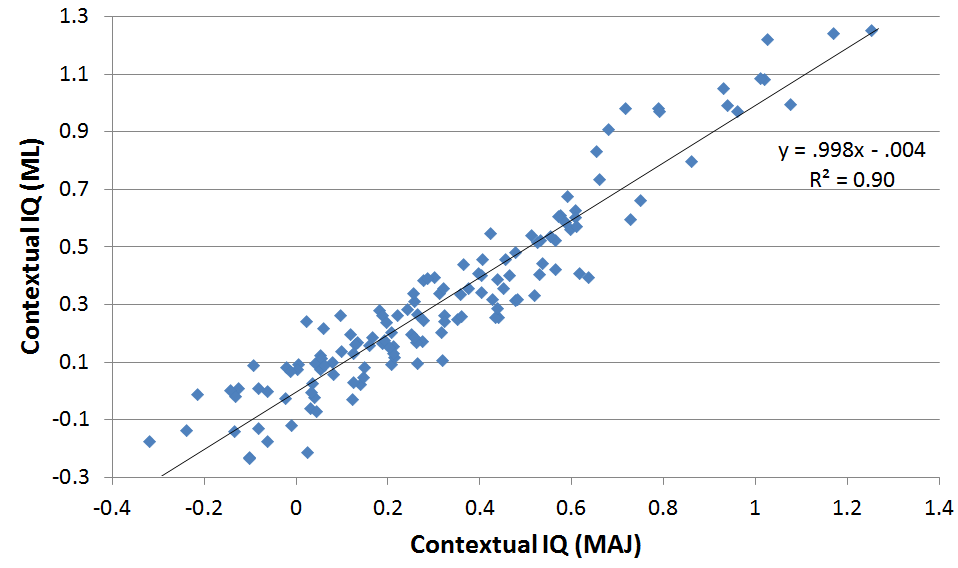}
 \caption{Contextual IQ under the majority and machine learning aggregator}
 \label{l_fig_ContextualIQ-MAJ-vs-ML}
\end{figure}

We now examine the extent to which a participant's contextual IQ is sensitive to the aggregator. Figure~\ref{l_fig_ContextualIQ-MAJ-vs-ML} shows a plot correlating a participant's Contextual IQ under the MAJ aggregator and her Contextual IQ under the ML aggregator.

Figure~\ref{l_fig_ContextualIQ-MAJ-vs-ML} shows a high correlation between participant's contextual IQ under the MAJ and ML aggregators (correlation coefficient of over 0.95). Thus, although the aggregator has a slight impact on contextual IQ, the key factors affecting contextual IQ are the participant's IQ and the participant match with the crowd (i.e. the uniqueness of her contribution).

\section{Conclusions and Limitations}
\label{l_conc}

In this paper we focused on measuring the quality of decisions based on aggregated opinions of the crowd. We proposed that the aggregation of crowd opinions can be conveniently studied using the samples of individual responses to standardized ability tests, such as {R}aven's Standard Progressive Matrices. One of the main advantages of such samples is the ability to quantify both individual and crowd performance on the same scale. 

We showed that decisions based on the aggregated opinions of the crowd are of higher quality than the average quality of the individual member's opinions. A crowd of 14 individuals has an average IQ score of around 115, one standard deviation above the average individual score. This finding is especially important for crowdsourcing environments where it is hard or impossible to detect highly performing individuals prior to the decision making process. We showed that the decisions based on the aggregated opinions of homogeneous crowds are better than the decisions based on the crowds' best performing members, whereas the best approach for a heterogeneous population is to identify the best performing individual and base the decision on her opinions.
Our findings indicate that while an individual expert can be smarter than the general opinion pool, she cannot compete against the crowd of her highly performing colleagues, even if she outsmarts each of them individually.


Finally, we proposed the concept of contextual IQ that allows measuring individual contributions towards the aggregate IQ of the crowd. We showed that although the contribution is typically higher when the individual's IQ is higher, it also depends on the uniqueness of individual's contribution in the context of a given crowd.


{\bf Limitations:} Our approach has several limitations. First, in our setting the crowd members expressed their opinions independently. Such a situation is typical for many crowdsourcing environments, but our findings may not be relevant to contexts in which crowd members can discuss or compare their opinions. Second, we did not collect our data in an actual crowdsourcing environment, where the structure of the individual's opinion might be different from what we observed in our sample. For instance, while the dominant strategy for filling the SPM IQ test is to attempt to answer the item even if the correct response is unknown to the individual, in some crowdsourcing environments (e.g. Amazon Mechanical Turk) individuals may be punished for providing incorrect responses, and thus usually refrain from doing so. Finally, the performance of the ML model was not significantly or consistently higher than of the simple MAJ aggregator which suggests that there is a field for improvement. For example, a more advanced model could allow for non-uniform distribution of incorrect responses. 

Many questions are open for future research. Are there better aggregators that give a stronger boost to Crowd IQ? Which aggregators are better fitted for large and small crowds? Do such aggregation effects also occur in domains other than IQ and real-life crowdsourcing settings? Specifically, would aggregating responses in crowdsourcing settings, such as Amazon's Mechanical Turk, yield similar results? 
Can the match between an individual and a crowd be predicted using features such as personality, gender or country of origin? Can contextual IQ be efficiently used to select small crowds that would have a high performance in real-world tasks?

\bibliographystyle{plain}
\bibliography{AIQ2}

\begin{thebibliography}{10}

\bibitem{austen1996information}
D.~Austen-Smith and J.S. Banks.
\newblock Information aggregation, rationality, and the condorcet jury theorem.
\newblock {\em American Political Science Review}, pages 34--45, 1996.

\bibitem{aziz2009power}
H.~Aziz, O.~Lachish, M.~Paterson, and R.~Savani.
\newblock Power indices in spanning connectivity games.
\newblock {\em Algorithmic Aspects in Information and Management}, pages
  55--67, 2009.

\bibitem{bachrach2010approximating}
Y.~Bachrach, E.~Markakis, E.~Resnick, A.D. Procaccia, J.S. Rosenschein, and
  A.~Saberi.
\newblock Approximating power indices: theoretical and empirical analysis.
\newblock {\em Autonomous Agents and Multiagent Systems}, 2010.

\bibitem{bachrach2009power}
Y.~Bachrach and J.S. Rosenschein.
\newblock Power in threshold network flow games.
\newblock {\em Autonomous Agents and Multi-Agent Systems}, 18(1):106--132,
  2009.

\bibitem{bachrach2008power}
Y.~Bachrach, J.S. Rosenschein, and E.~Porat.
\newblock Power and stability in connectivity games.
\newblock In {\em Proceedings of the 7th international joint conference on
  Autonomous agents and multiagent systems-Volume 2}, pages 999--1006.
  International Foundation for Autonomous Agents and Multiagent Systems, 2008.

\bibitem{banzhaf1965}
J.~F. Banzhaf.
\newblock Weighted voting doesn't work: a mathematical analysis.
\newblock {\em Rutgers Law Review}, 19:317--343, 1965.

\bibitem{dawid1979maximum}
A.P. Dawid and A.M. Skene.
\newblock Maximum likelihood estimation of observer error-rates using the em
  algorithm.
\newblock {\em Applied Statistics}, pages 20--28, 1979.

\bibitem{dekel2008incentive}
O.~Dekel, F.~Fischer, and A.D. Procaccia.
\newblock Incentive compatible regression learning.
\newblock In {\em SODA}, 2008.

\bibitem{everaere2007strategy}
P.~Everaere, S.~Konieczny, and P.~Marquis.
\newblock The strategy-proofness landscape of merging.
\newblock {\em Journal of Artificial Intelligence Research}, 2007.

\bibitem{gottfredson1997g}
L.S. Gottfredson.
\newblock Why g matters: The complexity of everyday life.
\newblock {\em Intelligence}, 24(1):79--132, 1997.

\bibitem{jensen1998g}
A.R. Jensen.
\newblock The g factor: The science of mental ability.
\newblock {\em London: Westport}, 1998.

\bibitem{kasneci2010bayesian}
G.~Kasneci, J.~Van~Gael, R.~Herbrich, and T.~Graepel.
\newblock Bayesian knowledge corroboration with logical rules and user
  feedback.
\newblock In {\em Proceedings of the 2010 European conference on Machine
  learning and knowledge discovery in databases: Part II}, pages 1--18.
  Springer, 2010.

\bibitem{Koller+Friedman:09}
D.~Koller and N.~Friedman.
\newblock {\em Probabilistic Graphical Models: Principles and Techniques}.
\newblock MIT Press, 2009.

\bibitem{Leech:2002:IMF}
Dennis L.
\newblock Voting power in the governance of the international monetary fund.
\newblock {\em Annals of Operations Research}, 109(1-4):375--397, 2002.

\bibitem{list2001epistemic}
C.~List and R.E. Goodin.
\newblock Epistemic democracy: generalizing the condorcet jury theorem.
\newblock {\em Journal of Political Philosophy}, 9(3):277--306, 2001.

\bibitem{list2009judgment}
C.~List and C.~Puppe.
\newblock Judgment aggregation: A survey.
\newblock {\em Handbook of Rational and Social Choice}, 2009.

\bibitem{lorenz2011social}
J.~Lorenz, H.~Rauhut, F.~Schweitzer, and D.~Helbing.
\newblock How social influence can undermine the wisdom of crowd effect.
\newblock {\em Proceedings of the National Academy of Sciences}, 2011.

\bibitem{lubinski2004introduction}
D.~Lubinski.
\newblock Introduction to the special section on cognitive abilities: 100 years
  after spearman's (1904)''general intelligence,'objectively determined and
  measured".
\newblock {\em Journal of Personality and Social Psychology}, 86(1):96, 2004.

\bibitem{lyle2008collective}
J.A. Lyle.
\newblock Collective problem solving: Are the many smarter than the few?
\newblock 2008.

\bibitem{mclennan1998consequences}
A.~McLennan.
\newblock Consequences of the condorcet jury theorem for beneficial information
  aggregation by rational agents.
\newblock {\em American Political Science Review}, pages 413--418, 1998.

\bibitem{Pearl:88}
J.~Pearl.
\newblock {\em Probabilistic reasoning in intelligent systems : networks of
  plausible inference}.
\newblock 1988.

\bibitem{pennock2007computational}
D.M. Pennock and R.~Sami.
\newblock Computational aspects of prediction markets, 2007.

\bibitem{raven:spm:2008}
J.C. Raven.
\newblock Standard progressive matrices plus, sets a-e.

\bibitem{raven1938progressive}
J.C. Raven.
\newblock {\em Progressive matrices}.
\newblock {\'E}ditions Scientifiques et Psychotechniques, 1938.

\bibitem{raven2000raven}
J.C. Raven.
\newblock The raven's progressive matrices: Change and stability over culture
  and time.
\newblock {\em Cognitive Psychology}, 41(1):1--48, 2000.

\bibitem{raven1978manual}
J.C. Raven, J.H. Court, and J.E. Raven.
\newblock {\em Manual for Raven's progressive matrices and vocabulary scales}.
\newblock HK Lewis, 1978.

\bibitem{samet1984application}
D.~Samet, Y.~Tauman, and I.~Zang.
\newblock An application of the aumann-shapley prices for cost allocation in
  transportation problems.
\newblock {\em Mathematics of Operations Research}, pages 25--42, 1984.

\bibitem{schmidt2004general}
F.L. Schmidt and J.~Hunter.
\newblock General mental ability in the world of work: occupational attainment
  and job performance.
\newblock {\em Journal of Personality and Social Psychology}, 86(1):162, 2004.

\bibitem{sen1986social}
A.~Sen.
\newblock Social choice theory.
\newblock {\em Handbook of mathematical economics}, 3:1073--1181, 1986.

\bibitem{shapley53value}
L.~S. Shapley.
\newblock A value for n-person games.
\newblock {\em Contrib. to the Theory of Games}, pages 31--40, 1953.

\bibitem{shapley_shubik1954}
L.~S. Shapley and M.~Shubik.
\newblock A method for evaluating the distribution of power in a committee
  system.
\newblock {\em American Political Science Review}, 48:787--792, 1954.

\bibitem{spearman1927abilities}
C.~Spearman.
\newblock The abilities of man.
\newblock 1927.

\bibitem{tijs1986game}
S.H. Tijs and T.S.H. Driessen.
\newblock Game theory and cost allocation problems.
\newblock {\em Management Science}, pages 1015--1028, 1986.

\bibitem{welinder2010multidimensional}
P.~Welinder, S.~Branson, S.~Belongie, and P.~Perona.
\newblock The multidimensional wisdom of crowds.
\newblock In {\em Neural Information Processing Systems Conference (NIPS)},
  volume~6, page~8, 2010.

\bibitem{woolley2010evidence}
A.W. Woolley, C.F. Chabris, A.~Pentland, N.~Hashmi, and T.W. Malone.
\newblock Evidence for a collective intelligence factor in the performance of
  human groups.
\newblock {\em Science}, 330(6004):686, 2010.

\end{thebibliography}
\end{document}